\documentclass[journal]{IEEEtran}

\ifCLASSINFOpdf
   \usepackage[pdftex]{graphicx}
   \DeclareGraphicsExtensions{.pdf,.jpeg,.png}
\else
   \usepackage[dvips]{graphicx}
   \DeclareGraphicsExtensions{.eps}
\fi
\usepackage{amssymb}
\usepackage{amsbsy}
\usepackage{amsmath}
\usepackage{amsfonts}

\usepackage{array}

\usepackage{mdwmath}
\usepackage{mdwtab}

\ifCLASSOPTIONcaptionsoff
  \usepackage[nomarkers]{endfloat}
 \let\MYoriglatexcaption\caption
 \renewcommand{\caption}[2][\relax]{\MYoriglatexcaption[#2]{#2}}
\fi
\begin{document}
%
\title{Cyclostationary Approach for Heart and Respiration Rates
Monitoring with Body Movement Cancellation Using Radar Doppler
System}
%
%
%
\author{Somayeh Kazemi, Ayaz Ghorbani, Hamidreza Amindavar and Changzhi Li., \emph{Member, IEEE} 
\thanks{Somayeh Kazemi, Ayaz Ghorbani and HamidReza Amindavar are with the
Department of Electrical Engineering, University of Amir Kabir,
Tehran, Iran. E-Mail: s.kazemi@aut.ac.ir, ghorbani@aut.ac.ir, hamidami@aut.ac.ir
 
Changzhi Li. is with the Department of Electrical and Computer Engineering, Texas Tech University, Lubbock, TX 79409 USA (e-mail: changzhi.li@ttu.edu).}
}

\maketitle

\begin{abstract}
Heart and respiration rate measurement using Doppler radar
is a non-contact and non-obstructive way for remote thorough-clothing
monitoring of vital signs. The modulated back-scattered radar signal in the presence of high noise and interference is non-stationary with hidden periodicities, which cannot be detected by ordinary Fourier analysis. In this paper we propose a cyclostationary approach for such signals and show that by using non-linear transformation and then Fourier analysis of the radar signal, the hidden periodicities can be accurately obtained. Numerical results show that the vital signs can be extracted as cyclic frequencies, independent of
SNR and without any filtering or phase unwrapping.
\end{abstract}


%
\IEEEpeerreviewmaketitle

\section{Introduction}
%
%
%
%
\IEEEPARstart{F}{or} a long time, non-contact monitoring of
heart and respiratory activity of patients and elderly
persons, based on CW microwave Doppler radar phase modulation, has been a valuable approach to remote sensing diagnosis of
syndromes risks and heart attacks in medicine and telemedicine
\cite{1}-\cite{3}. Many applications, such as baby monitoring to prevent sleep apnea or vital sign detection through clothing,  have motivated several studies to find the best methods to extract information from a signal modulated
by small chest motion \cite{4,5}. The main challenges in
these studies pivot around accurate information extraction in the
presence of high noise, clutter, and body motion interference
\cite{6,7}. A quadrature receiver is the best approach to prevent phase demodulation null points by choosing the
larger of the two signals, through direct phase demodulation or by
combining the signals \cite{8,9}.
The Gram-Schmidt technique for known
phase and amplitude imbalances is used to orthonormalize the two quadrature vectors
and analyze the error \cite{10}. Equal-ratio combining,
maximum-ratio combining, and principal component analysis also are
some data-driven approaches that are used to combine two
channel signals \cite{11}. Although high RF frequencies (small wavelengthes) result in high signal-to-noise ratio (SNR) of the detected chest-wall motion signal, the small angle approximation then
does not hold, and, hence, nonlinear combining techniques
are needed \cite{12,13}. The Levenberg-Marquardt (LM) estimation algorithm is a nonlinear method used to center the I-Q arc and to remove dc offsets and imbalances \cite{14,15}. Although successful Doppler radar non-contact vital
sign detection under different environments for stationary persons
has been reported in recent years \cite{2,3}, \cite{16}-\cite{18}, a
number of challenges still remain, including
development of nonlinear channel combining algorithms \cite{12,13},
cancelation of motion artifacts and background clutter
\cite{12}, \cite{19}, and  improvement of rate estimation methods under low SNR conditions \cite{20}.

As a matter of fact, since non-contact vital sign estimation is
based on detection of phase modulation induced by small chest motion in the millimeter range, the presence of environmental noise and random body
movements that occur in situations with small
transceivers and large distances are difficult challenges for obtaining
accurate results.

In this paper, we present a non-stationary approach to
vital sign detection and characterization in  the presence of noise and interference.
Many common statistical signal processing methods treat random
signals as if they were statistically stationary, but generally the
parameters of the underlying physical process that generates the signals vary with time. The detection, analysis, and
feature extraction of signals involving general unknown
non-stationarities, with only a few signal records, is generally impossible.
Fortunately, some real-world signals such as heart and respiration
modulated signals have some parameters that vary periodically or
almost periodically with time. This leads to the fact that a
random signal with property periodicity could be modeled as a
cyclostationary process \cite{21,22}.

Cyclostationary theory is one of the most suitable methods for
analyzing signals, that have a cyclic pattern of statistical
properties. It mainly uses the cyclic statistics to analyze, detect,
and estimate signals with hidden periodicities.
Another basic advantage is the
robustness of cyclostationary processing in an environment with
high noise and interference \cite{23}-\cite{26}. For these reasons, the
cyclostationaity approach can be utilized in monitoring heart and respiratory activity of patients, as a robust
signal processing method, and to obtain accurate rate estimation using
a phase modulated signal in situations with very low SNR and in
the presence of two-dimensional (2-D) random body movements, for which common filtering approaches fail. The analytical development of the cyclostationairty approach is straight-forward
for accurately extracting the hidden periodicities in the signal returned from
a patient, where existing common frequency
demodulation methods and Fourier analysis would not work in the presence of high noise and body
motion interference \cite{27}-\cite{30}.

This paper is organized
as follows, Doppler radar  background for vital sign
monitoring is presented in Section II. In Sections III and IV, the
fundamental second-order cyclostationary formulation is presented for a time
series and statistical approach. In Section V,
simulated data is applied to validate this theory, and it is shown
that the method is insensitive to SNR level and therefore robust
to very low SNR and notable random body motion.



\section{Radar Doppler Background}

A Doppler radar  vital sign sensing transceiver transmits a CW radio
signal and receives a motion-modulated signal reflected from
the moving chest wall of the subject. Two main periodic motions due to heart beat and
respiration affect the chest motion. These two motions can be modeled as, respectively,
\begin{eqnarray}
h(t)=a_h \cos(2\pi f_h t)
\label{eq:a1}
\end{eqnarray} and
\begin{eqnarray}
r(t)=a_r \cos(2\pi f_r t)
\label{eq:a2}
\end{eqnarray} where parameters $a_h$, $a_r$, $f_h$, and $f_r$ are, respectively, the heart and respiration beat
component amplitudes and frequencies. Therefore, based on the Doppler effect, the RF wave reflected
from the surface of the patient's chest, undergoes two main phase
shifts proportional to the surface displacement due to these two components.

In  front of the radar system with sufficient
beam-width, the subject will typically have some random body movement in
the $x$ and $y$ directions. For full-dimentional
monitoring capability at all possible angles that a subject may
move, two identical radar systems with relatively large beam-width
antennas can be implemented at two corners of the room as shown in
Fig. 1. In each successive time interval, the larger of the receiver signals from the two radar systems is
processed. Slow movements in the
overall, practical, and least informative scenario can be modeled as
a one-dimensional (1-D) random process with uniform distribution in the
specified interval for both directions. Therefore, 2-D random motion of the subject for each
receiver is 1-D and the phase of the baseband signal for each system is affected by the sum of the 1-D random motions and
residual phase noise \cite{6}. Thus, we formulate the resulting in-phase (I) and quadrature (Q) components of the RF signal as,
\begin{eqnarray}
B_\mathrm{I} (t)&\!\!\!\!\!\!\!=\!\!\!\!\!\!\!&
A\cos\left\{\frac{4\pi}{\lambda}
\left[a_h \cos(2\pi f_h t)+a_r \cos(2\pi f_r
t)\right.\right.\nonumber\\
&&\vspace{2cm}+\left.\left.x(t)\right]+\phi_\mathrm{n}
(t)+C\rule{0pt}{14pt}\right\}+N_\mathrm{I}(t),\label{eq:a3}\\
B_\mathrm{Q} (t)&\!\!\!\!\!\!\!=\!\!\!\!\!\!\!&
A\sin\left\{\frac{4\pi}{\lambda}
\left[a_h \cos(2\pi f_h t)+a_r \cos(2\pi f_r
t)\right.\right.\nonumber\\
&&\vspace{2cm}+\left.\left.x(t)\right]+\phi_\mathrm{n}
(t)+C\rule{0pt}{14pt}\right\}+N_\mathrm{Q}(t).\label{eq:a4}
\end{eqnarray} where the signals $\phi_\mathrm{n} (t)$, $N_\mathrm{I} (t)$, and $N_\mathrm{Q} (t)$ denote the phase noise and
the receiver noises in the two  quadrature channels, respectively, $C$ represents the dc value of the initial subject range, and $\lambda$ denotes the wavelength of the electromagnetic probing signal. This modeling would account for all possible random body movements.
The main sources of phase noise $\phi_\mathrm{n} (t)$ and its statistically model
is discussed in Appendix A. The $N_\mathrm{I} (t)$ and $N_\mathrm{Q} (t)$ noise is
usually modeled as i.i.d. (identically independently distributed)
Gaussian white noise with zero mean and variance $\sigma^2$.

We define two parameters to simplify subsequent formulations,
\begin{eqnarray}
A_h=\frac{4\pi}{\lambda} a_h,\label{eq:a5}\\
A_r=\frac{4\pi}{\lambda} a_r.\label{eq:a6}
\end{eqnarray} Because the chest moves a greater distance over a greater area due to breathing than it
does for the heart beating, the amplitude of the
respiration signal is typically about 100 times greater than that of
the heart beat signal \cite{2}. The resting heart rate is
generally between 0.83 and 1.5 Hz (50 and 90 beats per minute),
while the resting respiration rate is generally between 0.15 and 0.4
Hz (9 and 24 breaths per minute) \cite{2}.

Because of the random body motion and noise components, (\ref{eq:a3}) and (\ref{eq:a4}) constitute
two random signals. A time-varying statistical mean of these signals is calculated as shown below,
\begin{eqnarray}
\mu_\mathrm{I}(t)&=&<B_\mathrm{I}(t)>=\lim_{T\rightarrow\infty}\frac{1}{T} \int_{\frac{t-T}{2}}^{\frac{t+T}{2}}\nonumber\\
&&\times\left\{A\cos[A_r \cos(2\pi f_r u)+A_h \cos(2\pi f_h u)\right.\nonumber\\
&&\left.+x(u)+\phi_n (u)+C]+N_\mathrm{I} (u)\right\}du \label{12}\\
\mu_\mathrm{Q} (t)&=&<B_\mathrm{Q} (t)>=\lim_{T\rightarrow\infty}\frac{1}{T} \int_{\frac{t-T}{2}}^{\frac{t+T}{2}}\nonumber\\
&&\times\left\{A\sin[A_r \cos(2\pi f_r u)+A_h \cos(2\pi f_h u)\right.\nonumber\\
&&\left.+x(u)+\phi_n (u)+C]+N_\mathrm{Q} (u)\right\}du\label{13}
\end{eqnarray} where $<.>$ denotes the time average operator. Thus, the received vital signals are inherently non-stationary
random processes, which have no obvious periodicity and hence no lines in the spectral density, so the information obtained
from ordinary frequency analysis methods is limited. In the next section, we prove
that with the deployment of cyclostationary theory, the rate and amplitude information can be extracted in high noise and interference,
without any demodulation, filtering, or phase unwrapping. The results conducted later on simulated data will demonstrate the effectiveness of the proposed method.

\begin{figure}[!t]
\centering
\includegraphics[width=2.5in]{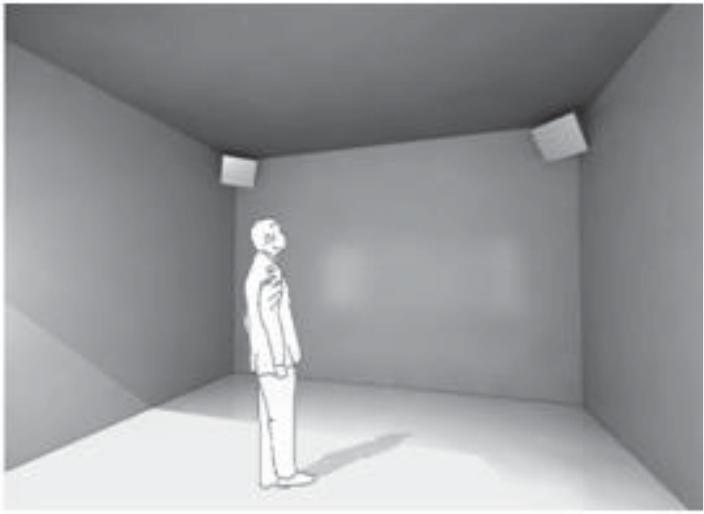}
\caption{Two radar systems at two sides of a room, where the 2-D motion
of the subject is 1-D for each system.}
\label{fig1}
\end{figure}

\section{Cyclostationary Analysis of Time Series Signal }

As noted, according to the Doppler effect, the reflected signal from a moving target has a modulated phase in proportion to the target position.  The goal is the extraction of the heart and
respiration periodic rates and amplitudes from
the received modulated random signal, with the addition of minimal
amounts of in-band noise. The existence of noise [$\phi_\mathrm{n} (t)$,
$N_\mathrm{I} (t)$, and $N_\mathrm{Q} (t)$] and interference
resulting from random body movement [$x(t)$] in (\ref{eq:a3}) and (\ref{eq:a4}),
in addition to the two periodic components (heart and respiration), cause non-stationary behavior of the signal that
involves two hidden periodicities. By definition, the
cyclostationary signal has a random statistically non-stationary nature, with periodic behavior in one or more of its
parameters  \cite{21}. In order to show the basic cyclic structure
of the received radar signals and introduce the cyclostationarity approach, we apply the analytical
form of the I-Q signals obtained from the two receiver chains. It is noted that
two quadrature receiver chains are used to prevent the null point
problem  \cite{2}. Accordingly, the analytic signal is formulated as,
\begin{eqnarray}
\label{eq:a9}
y(t)&=&B_\mathrm{I}(t)+jB_\mathrm{Q} (t)\nonumber\\
&=& D\exp[jA_h  \cos(2\pi f_h t)]\nonumber\\
&&\times\exp\left[jA_r  \cos(2\pi f_r t) \right]\exp\left[j \frac{4\pi}{\lambda} x(t) \right]\nonumber\\
&&\times\exp\left[j\phi_\mathrm{n} (t) \right]+N_\mathrm{I} (t)+jN_\mathrm{Q} (t)
\end{eqnarray} where, \begin{equation} D=A\exp(jC). \nonumber \end{equation} We re-express (\ref{eq:a9}) as
\begin{eqnarray}
y(t)&=& D m(t)\exp[jA_h\cos(2\pi f_h t)]\nonumber\\&&\times\exp[jA_r\cos(2\pi f_r t)]+z(t),\nonumber\\
\label{eq:a10}
\end{eqnarray} where, \begin{eqnarray} z(t)=N_\mathrm{I} (t)+jN_\mathrm{Q} (t) \nonumber \end{eqnarray} and
\begin{eqnarray}
m(t)=\exp\left[j\frac{4\pi}{\lambda} x(t)\right]\exp\left[j\phi_\mathrm{n}(t)\right],
\label{eq:a11}
\end{eqnarray} is the exponential of the random motion and phase noise, where $x(t)$ and $\phi_\mathrm{n}(t)$ are assumed real and mutually independent.
As noted, the noises $N_\mathrm{I} (t)$ and $N_\mathrm{Q} (t)$ are zero-mean white Gaussian. The signal $y(t)$ in (\ref{eq:a10}) is in a basic cyclic form that has cyclostationary properties and can be interpreted as an amplitude and frequency modulated (AM-FM) signal \cite{31}. The periodic signal
$\exp[jA_h\cos(2\pi f_h t)]$ can be thought of as a phase
modulated signal with modulating signal $A_h\cos(2\pi f_ht)$, where the
RF signal (before modulation) has been obtained by downshifting the original RF signal, and likewise for the signal $\exp[jA_r\cos(2\pi f_r
t)]$. Finally, since frequency modulation is the derivative of phase modulation, each of the signals
\begin{eqnarray}
s_h(t)=\exp[jA_h\cos(2\pi f_h t)],
\label{eq:b11}
\end{eqnarray} and
\begin{eqnarray}
s_r(t)=\exp[jA_r\cos(2\pi f_r t)],
\label{eq:b12}
\end{eqnarray} is therefore a complex baseband signal representing a real FM signal with a single tone
as the modulating signal. The multiplication of these two signals is
AM modulated by signal $m(t)$. Therefore, $y(t)$ is an AM-FM
modulated signal with amplitude $m(t)$  and multi-carrier frequency. Signals
$s_h (t)$ and $s_r (t)$ without any nonlinear transformation have spectral lines
and therefore, would be first-order
cyclostationary processes. The frequency-domain representations of $s_h(t)$ and $s_r(t)$ are given by \cite{21}

\begin{eqnarray}
S_h (f)&=&\sum_{n=-\infty}^{\infty}J_n (A_h)\delta(f-nf_h)
\label{eq:a12},\\
S_r (f)&=&\sum_{n=-\infty}^{\infty}J_n (A_r)\delta(f-nf_r). \label{eq:a13}
\end{eqnarray} where $J_n$ is the first order Bessel function of order $n$. Thus, the periodic signals
$s_h(t)$ and $s_r(t)$ nominally contain all harmonics of $f_h$ and $f_r$, with specific
magnitudes given by the Bessel functions evaluated at $A_h$ and $A_r$, respectively.
The carrier frequencies of the multi-carrier AM-FM signal $y(t)$  can
be determined by convolving the impulsive spectra $S_h(f)$ and $S_r(f)$. The resultant discrete spectrum
contains all sums and differences of the frequencies of the original
two discrete spectra; that is, all sums and differences of all
harmonics of $f_h$ and $f_r$. Thus we can write

\begin{eqnarray}
S_y (f)&=& DM(f)* \sum_{n=-\infty}^\infty\sum_{m=-\infty}^\infty J_n(A_h)J_m (A_r)\nonumber\\
&&\times\delta[f-(nf_h+mf_r)]+N_\mathrm{I}(f)+jN_\mathrm{Q}(f),
\label{eq:a14}
\end{eqnarray} where $M(f)$ is the Fourier transform of $m(t)$ in (\ref{eq:a11}). Since the signal $y(t)$ is contaminated by strong noise terms
[$\phi_\mathrm{n} (t)$, $N_\mathrm{I} (t)$, $N_\mathrm{Q} (t)$], random body motion
interference $[x(t)]$, and also the dc term ($C$), its periodicities would be
hidden, and so conventional frequency analysis methods are not suitable
to extract the hidden information.

In Fig. 2 we depict one realization of the random process $y(t)$ in (\ref{eq:a10}) for a carrier frequency of 2.4 GHz ($\lambda=$ 0.125 cm), where the $a_h$, $a_r$, $f_h$, and $f_r$ values in this example are assumed as 0.01 cm, 1 cm, 1.9 Hz, and 0.4 Hz, respectively. The variance of  zero-mean white Gaussian noises $N_\mathrm{I} (t)$ and $N_\mathrm{Q} (t)$ is assumed equal to 5. Phase noise $\phi_\mathrm{n} (t)$ is zero-mean Gaussian noise with variance 5.
Random body motion $x(t)$ is assumed a uniform random process between zero to 10 cm [$U\sim(0,10)$]. As shown in Fig. 2, there is no visible periodicity in the $y(t)$ signal.
\begin{figure}[!t]
\centering
\includegraphics[width=3.5in]{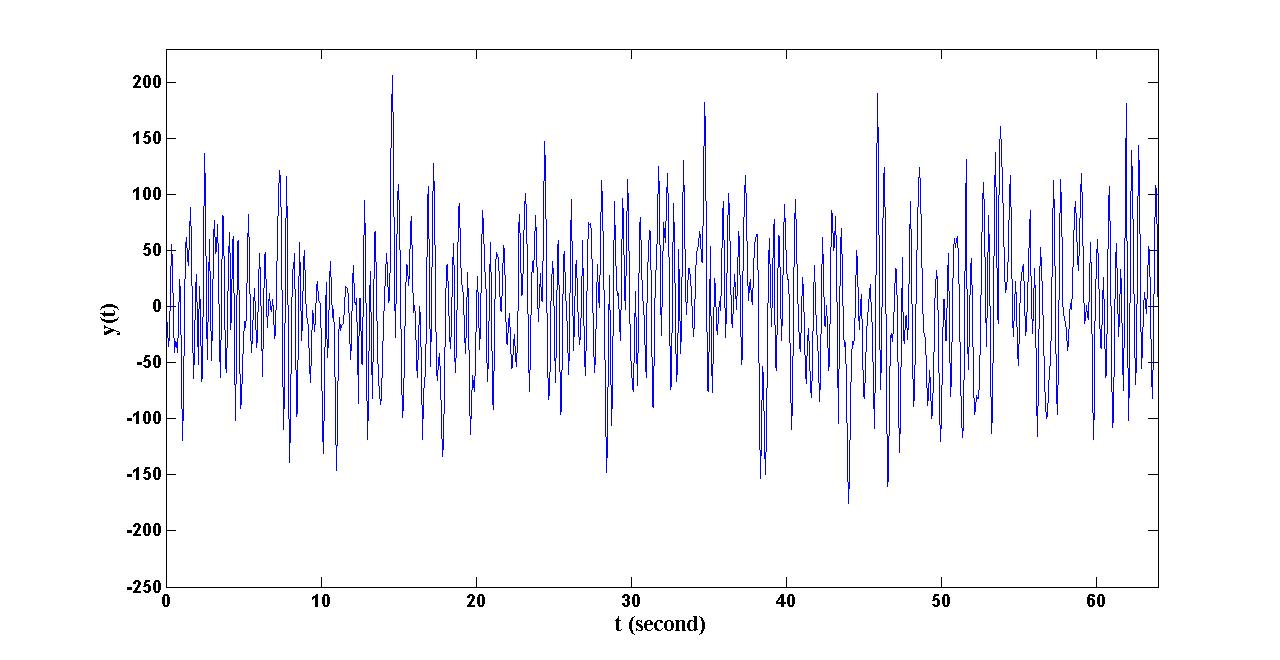}
\caption{One realization of the complex radar signal $y(t)$. }
\label{fig2}
\end{figure}

Now, we turn to higher-order frequency analysis, which concentrates on the
periodic features of the signal, and will prove that a signal with
structure (\ref{eq:a14}) is a cyclostationary signal. By definition, a signal is cyclostationary of
order \emph{$n$} if and only if  an \emph{$n$}th-order
nonlinear transformation of the signal can be found that generates finite
sine wave components \cite{21}. In this
section, we will prove that the signal $y(t)$ is cyclostationary of second-order
and the related nonlinear (quadrature)
transformation to generate sine wave components is the limit
periodic autocorrelation function \cite{32}. As is obvious from
(\ref{eq:a14}), each one of the spectral lines is of magnitude
\begin{eqnarray}
A(n,m)=J_n(A_h)J_m (A_r)
\label{eq:b13}
\end{eqnarray} and frequency $nf_h+mf_r$, being carriers of an AM signal
having baseband modulating signal $m(t)$. Therefore, the cycle
frequencies $\{\alpha\}$ of $y(t)$ are of the conjugate type and for
order 2, are equal to the second harmonics of all the multiple
carrier frequencies: $\alpha=2nf_h+2mf_r$. All higher-order cycle
frequencies are simply sums and differences of these 2nd-order cycle
frequencies, and contain no new information about the
cyclostationarity of this signal. Therefore, this signal has two
periodic components and so is called second-order cyclostationary
\cite{21}. Accordingly, any attempt to apply the cyclostationarity approach should use the second-order cyclostationarity properties, i.e., the instantaneous conjugate autocorrelation or conjugate spectral
correlation functions. The limit periodic autocorrelation functions
of signals $s_h(t)$ and $s_r(t)$  are expressed as \cite{32}:
\begin{eqnarray}
R_{s_h}(t,\tau)&=&\lim_{N\to\infty}\frac{1}{2N+1}\nonumber\\
&&\sum_{n=-N}^N s_h\left(t+nT_h+\frac{\tau}{2}\right)s_h^*\left(t+nT_h-\frac{\tau}{2}\right)\nonumber\\
&=&\sum_{n=-\infty}^\infty\sum_{m=-\infty}^\infty J_n(A_h)J_m(A_r)\nonumber\\
&&\times e^{j2\pi f_h \left[m\left(t+\frac{\tau}{2}\right)-n\left(t-\frac{\tau}{2}\right)\right]},\label{eq:a15}
\end{eqnarray}
\begin{eqnarray}
R_{s_r}(t,\tau)&=&\lim_{N\to\infty}\frac{1}{2N+1}\nonumber\\
&&\sum_{n=-N}^N s_r\left(t+nT_r+\frac{\tau}{2}\right)s_r^*\left(t+nT_r-\frac{\tau}{2}\right)\nonumber\\
&=&\sum_{n=-\infty}^\infty\sum_{m=-\infty}^\infty J_n  (A_h ) J_m(A_r )\nonumber\\
&&\times e^{j2\pi f_h\left[m\left(t+\frac{\tau}{2}\right)-n\left(t-\frac{\tau}{2}\right)\right]}
\label{eq:a16}
\end{eqnarray} where $T_h=1/f_h$  and $T_r=1/f_r$. In
Appendix A, it is proved that the autocorrelation function of signal $m(t)$
is a function of the lag index only. Therefore, the overall
limit periodic autocorrelation function is written,
\begin{eqnarray}
R_y(t,\tau)&=&\lim_{N\to\infty}\frac{1}{2N+1} \sum_{n=-N}^N y\left(t+\frac{\tau}{2}\right)y^*\left(t-\frac{\tau}{2}\right)\nonumber\\
&=&DR_m(\tau) R_{s_h} (t,\tau) R_{s_r}(t,\tau)\nonumber\\
&&+R_{N_\mathrm{I}}(\tau)+jR_{N_\mathrm{Q}}(\tau)\nonumber\\
&=&DR_m (\tau)\sum_{k=-\infty}^\infty\sum_{l=-\infty}^\infty
\sum_{m=-\infty}^\infty\sum_{n=-\infty}^\infty \nonumber\\
&&\times J_k(A_h )J_l(A_h )J_m(A_r )J_n(A_r )\nonumber\\
&&\times e^{j2\pi f_h[k\left(t+\frac{\tau}{2}\right)-l\left(t-\frac{\tau}{2}\right)]}\nonumber\\
&&\times e^{j2\pi f_r[m\left(t+\frac{\tau}{2}\right)-n\left(t-\frac{\tau}{2}\right)]}\nonumber\\
&&+R_{N_\mathrm{I}}(\tau)+jR_{N_\mathrm{Q}}(\tau).
\label{eq:a17}
\end{eqnarray} Since the limit periodic autocorrelation function of the signal
$y(t)$ has lines in its spectral analysis, the signal
$y(t)$ is almost second-order cyclostationary with frequencies $f_h$
and $f_r$  \cite{21}. Now to seek the hidden
frequencies, it is helpful to localize the correlation of
frequency-shifted signals for a cyclostationary random signal $y(t)$
in the frequency domain. The cyclic-auto-correlation function of
$y(t)$ is the Fourier coefficient of the limit periodic
autocorrelation function at cyclo-frequency $\alpha$. Therefore, the
overall cyclic autocorrelation function is expressed,
\begin{eqnarray}
R_y^\alpha(\tau)&=&\int_{-\infty}^\infty R_y (t,\tau) e^{-j2\pi \alpha t} dt\nonumber\\
&=&DR_m (\tau)\sum_{k=-\infty}^\infty\sum_{l=-\infty}^\infty
\sum_{m=-\infty}^\infty\sum_{n=-\infty}^\infty \nonumber\\
&&\times J_k(A_h )J_l(A_h )J_m(A_r )J_n(A_r )\nonumber\\
&&\times e^{j\pi\left[\left(k+l\right)f_h+\left(m+n\right)f_r\right]}\nonumber\\
&&\times\delta\left\{\alpha-\left[(k-l) f_h+(m-n)f_r\right]\right\}\nonumber\\
&&+R_{N_\mathrm{I}}(\tau)+jR_{N_\mathrm{Q}}(\tau).
\label{eq:a18}
\end{eqnarray} Since the noises $N_\mathrm{I} (t)$ and $N_\mathrm{Q} (t)$ are assumed white,
the autocorrelation functions of $R_{N_\mathrm{I}}(t)$ and $R_{N_\mathrm{Q}}(t)$
are delta functions. Moreover, the spectral correlation function
(SCF) of the second-order cyclostationary signal $y(t)$  is
the Fourier transform of (\ref{eq:a18}), and is expressed as \cite{21}
\begin{eqnarray}
S_y^\alpha(f)&=&\int_{-\infty}^\infty R_y^\alpha(\tau) e^{-j2\pi\alpha\tau} d\tau\nonumber\\
&=&D\Re_m (f)\sum_{k=-\infty}^\infty\sum_{l=-\infty}^\infty
\sum_{m=-\infty}^\infty\sum_{n=-\infty}^\infty \nonumber\\
&&\times J_k(A_h)J_l(A_h)J_m(A_r)J_n(A_r)\nonumber\\
&&\times e^{j\pi\left[\left(k+l\right)f_h+\left(m+n\right)f_r\right]}\nonumber\\
&&\times\delta\left\{\alpha-\left[(k-l) f_r+(m-n)f_h\right]\right\}\nonumber\\
&&+C_{N_\mathrm{I}}+jC_{N_\mathrm{Q}}.
\label{eq:a19}
\end{eqnarray} According to (\ref{eq:a18}), the $R_y^\alpha (\tau)$ function involves
all frequencies and, therefore, the SCF is a 2-D function
that has all of the frequencies along one axis, showing discrete hidden frequencies and harmonics, and cyclic frequency along the other axis. The $\Re_m (f)$
function, which is the Fourier transform of the motion and phase
noise, acts as multiplicative noise for the delta functions, and terms of
$C_{N_\mathrm{I}}$ and $C_{N_\mathrm{Q}}$ which are the Fourier transforms of $R_{N_\mathrm{I}}(\tau)$ and $R_{N_\mathrm{Q}}(\tau)$ respectively, are two constants that don't affect the delta functions except at very low SNR. Moreover the SCF of the
cyclostationary signal along the cyclic axis is insensitive to all of
the noncyclic components. Cyclostationary
theory can be used as a powerful tool to extract the cyclic
frequencies from noisy non-stationary signals \emph{independent of the SNR level}.

\section{Statistical Cyclostationary Analysis}

In Section III, the second-order cyclostationary property of the
received radar signal was proven in a time series analysis. Now, a
statistical approach is presented to prove the cyclostationary
property for an ensemble in terms of expected value. According
to \cite{32}, the joint fraction-of-time amplitude distribution for
a time series $y(t)$ is defined by,
\begin{eqnarray}
F_{y(t_1)y^*(t_2)}(x_1,x_2)&\triangleq&\lim_{N\to\infty}\frac{1}{2N+1}\sum_{n=-N}^N \nonumber\\
&& U[x_1-y(t_1+nT_0)]\nonumber\\
&&\times U[x_2-y^*(t_2+nT_0)]\label{eq:a20}
\end{eqnarray} where  $U$ is the unit-step function and $T_0$ is the least common
multiple of $T_h$ and $T_r$. So the joint fraction-of-time amplitude density
for typical $y(t)$ is expressed as,
\begin{eqnarray}
f_{y(t_1)y^*(t_2)}(x_1, x_2)=\frac{\partial^2}{\partial y_1\partial
y^*_2}F_{y(t_1)y^*(t_2)}(x_1, x_2) \label{eq:a21}
\end{eqnarray} where both $F_{y_1(t_1)y^*_2(t_2)}$ and $f_{y_1(t_1)y^*_2(t_2)}$ are
jointly periodic with respect to both time variables $t_1$ and $t_2$, over period $T_0$.
It is shown in \cite{32} that the probabilistic autocorrelation is given by,

\begin{eqnarray}
E\{y(t_1)y^*(t_2)\}&\triangleq& \int_{-\infty}^{\infty}\int_{-\infty}^{\infty} x_1x_2f_{y(t_1)y^*(t_2)}(x_1, x_2)dx_1dx_2\nonumber\\
&&=\lim_{N\to\infty}\frac{1}{2N+1}\sum_{n=-N}^N \nonumber\\
&& y(t_1+nT_0)y^*(t_2+nT_0). \label{eq:a22}
\end{eqnarray} Therefore, from this and (\ref{eq:a17}), the limit period
autocorrelation can be interpreted as the probabilistic
autocorrelation,
\begin{eqnarray}
&&\hat R_y(t,\tau)=E\left[y(t+\frac{\tau}{2})y^*(t-\frac{\tau}{2})\right].
\label{eq:a24}
\end{eqnarray} Moreover, this expected value can be interpreted as an ensemble
average \cite{32}. According to the limit periodic autocorrelation function in
(\ref{eq:a17}) and its Fourier transforms in (\ref{eq:a18}) and (\ref{eq:a19}), it is proven that
the heart and respiration frequencies can be extracted independent of SNR level using the statistical approach.

Since the peak-to-peak chest motion due to heart beat is about 0.5 mm  whereas that due to respiration ranges from
4 mm to 12 mm \cite{2}, the values of $A_h$ and $A_r$ in (\ref{eq:a5}) and (\ref{eq:a6})
at a nominal RF frequency of 2.4 GHz would be 0.0502 and in the range of 0.401 to
1.205, respectively. Out of all possible combinations
over the respective ratios of spectral lines (\ref{eq:a19}), we search for the highest values
of $J_k (A_h)J_l (A_h)J_m (A_r)J_n (A_r)$ over these ranges
of the $A_r$ and $A_h$ parameters. It can be shown that if the relation of heart and respiration amplitudes is assumed as noted above, i.e
$\frac{1}{50}<\frac{a_h}{a_r}<\frac{1}{10}$, then the largest values
of $J_k(A_h)J_l(A_h)J_m(A_r)J_n(A_r)$ and respective frequencies are
ordered as shown in Table 1. Therefore, the heart amplitude
can be found from $\frac{b_6}{b_1}$ and $\frac{J_1(A_h)}{J_0(A_h)}$ relations and the respiration amplitude obtained from $\frac{b_1}{b_2}$
and $\frac{J_1 (A_r)}{J_0 (A_r)}$ relations.

\begin{table}[h]
\renewcommand{\arraystretch}{1.3}
\caption{Order of the largest values of $J_k (A_h)J_l
(A_h)J_m (A_r)J_n (A_r)$ and respective frequencies}
\centering
\begin{tabular}{|c||c||c|}
\hline
Order of peaks & Amplitude & Frequency\\
\hline
$b_1$ & $J_0 (A_h)J_0 (A_h)J_0 (A_r)J_0 (A_r)$ & 0\\
\hline
$b_2$ & $J_0 (A_h)J_0 (A_h)J_1 (A_r)J_0 (A_r)$ & $f_r$\\
\hline
$b_3$ & $J_0 (A_h)J_0 (A_h)J_2 (A_r)J_0 (A_r)$ & $2f_r$\\
\hline
$b_4$ & $J_0 (A_h)J_0 (A_h)J_3 (A_r)J_0 (A_r)$ & $3f_r$\\
\hline
$b_5$ & $J_0 (A_h)J_0 (A_h)J_4 (A_r)J_0 (A_r)$ & $4f_r$\\
\hline
$b_6$ & $J_1 (A_h)J_0 (A_h)J_0 (A_r)J_0 (A_r)$ & $f_h$\\
\hline
\end{tabular}
\end{table}

Simulation results will show that the hidden frequencies from the
signal contaminated by high noise and random body motion can
be accurately extracted  without any filtering, demodulation,
or phase unwrapping, and are also insensitive to SNR level.

\section{Simulation Results}

In this section, we demonstrate the robustness of the cyclostaionary
approach to vital-sign detection in heart and respiration monitoring for typical receiver signals in Doppler radar  systems. The spectral
correlation exhibited by cyclostationary or almost-cyclostationary
processes is completely characterized by the cyclic spectra
$S_y^\alpha (f)$ or equivalently by the cyclic auto-correlation $R_y^\alpha$ \cite{21}. In practice, the cyclic-spectral
density must be estimated because the signals being considered are
defined over a finite time interval $\Delta t$, and therefore, the
cyclic spectral density cannot be measured exactly. Estimates of the
cyclic spectral density are obtained via a time-smoothing technique
\cite{33}:
\begin{eqnarray}
&&S_y^\alpha(f)\approx  S_{y_{T_\mathrm{W}}}^\alpha (t,f)=\frac{\Delta t}{2}
\int^{t+\frac{\Delta t}{2}}_{t-\frac{\Delta t}{2}}
S_{y_{T_\mathrm{W}}}^\alpha(u,f) du \label{eq:a25}
\end{eqnarray}\\where
\begin{eqnarray}
&&S_{y_{T_\mathrm{W}}}^\alpha(u,f)=\frac{1}{T_\mathrm{W}}Y_{T_\mathrm{W}}
\left(u,f+\frac{\alpha}{2}\right)Y_{T_\mathrm{W}}^*
\left(u,f-\frac{\alpha}{2}\right) \label{eq:a26}
\end{eqnarray} $\Delta t$ is the total observation time of the signal, $T_\mathrm{W}$ is the short-time FFT window length, and
\begin{eqnarray}
Y_{T_\mathrm{W}}(t,f)=\int_{t-\frac{T_\mathrm{W}}{2}}^{t+\frac{T_\mathrm{W}}{2}}y(u)
e^{-j2\pi fu} du, \label{eq:a27}
\end{eqnarray}

Two examples with different SNR and SINR (signal-to-interference-plus-noise ratio) values will be simulated.
Here we define SNR as the ratio of (heart or respiration) signal
power to ($N_I$ or $N_Q$) receiver noise power, and the SINR as the ratio of (heart or
respiration) signal power to the sum of the powers of
motion interference $x(t)$ and phase noise $\phi_\mathrm{n}(t)$.
We assume that $x(t)$ is 1-D random motion with uniform distribution in the $x$
or $y$ direction, the sources of the phase noise $\phi_\mathrm{n}(t)$ are of
the type for which it has been proved that $\exp[j\phi_\mathrm{n}(t)]$ is stationary
(Appendix A), and $N_\mathrm{I} (t)$ and $N_\mathrm{Q} (t)$ are assumed i.i.d
Gaussian white noise with zero mean and variance $\sigma^2$. The first example is for the signal that is plotted in Fig. 2. With the parameter values assumed for this signal, the SNR and SINR are both equal to --30 dB for the heart signal and --10 dB for the respiration signal.

A plot of the power spectral density of this signal obtained, using an \emph{$n$}-point DFT MATLAB function with \emph{$n$} = 131072, is shown in Fig. 3. As can be seen, there is no information in the spectral density of the received signal about the heart and respiration frequencies and amplitudes.

\begin{figure}[!t]
\centering
\includegraphics[width=3.5in]{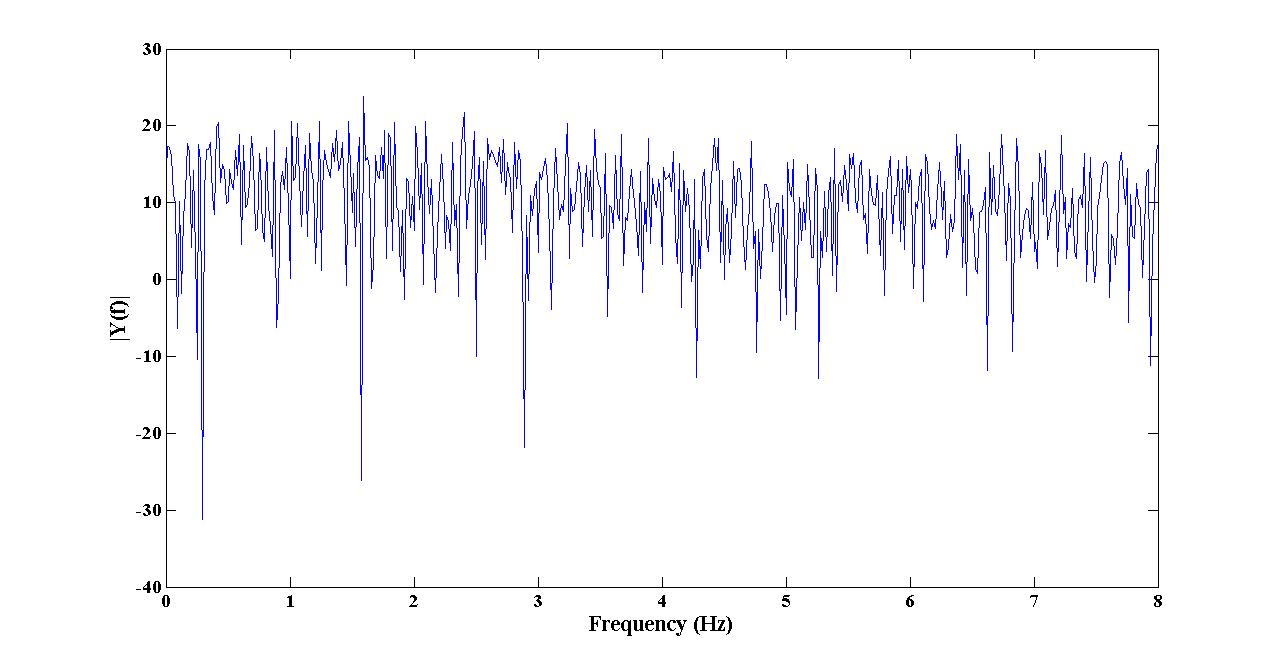}
\caption{Power spectral density of complex radar signal}
\label{fig4}
\end{figure}

\begin{figure}[!t]
\centering
\includegraphics[width=3.5in]{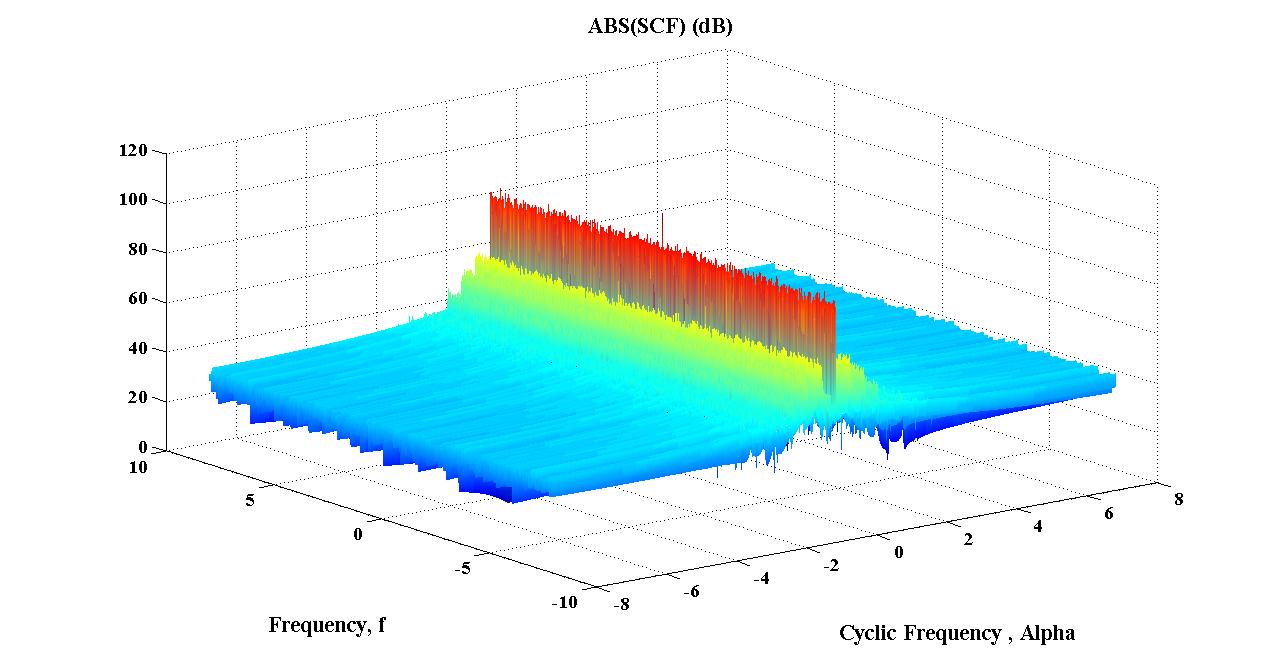}
\begin{center}
(a)
\end{center}
\includegraphics[width=3.5in]{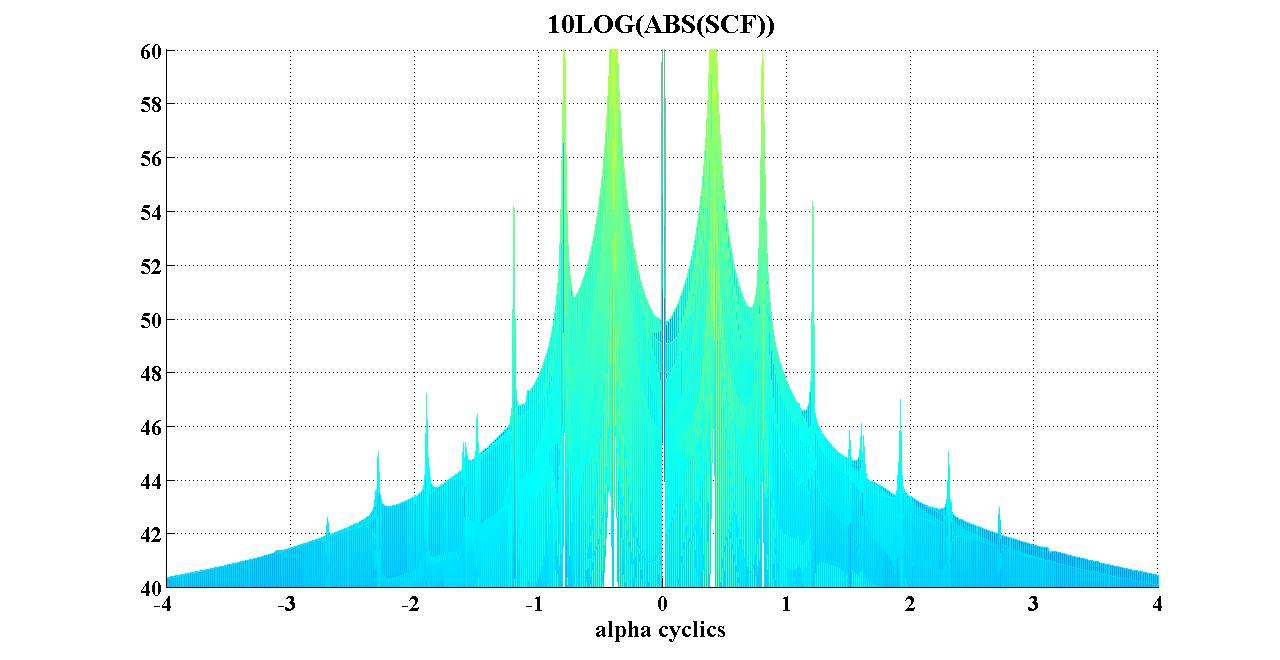}
\begin{center}
(b)
\end{center}
\caption{Spectral correlation function, (a) Overall view, (b)
Projection onto cyclic-frequency plane.}
\label{fig5}
\end{figure}


The SCF function (\ref{eq:a25}) of this signal is shown in Fig. 4 and
indicates that the signal has a continuous frequency distribution due
to the non-stationary nature at the frequency axis for $f$ and a
discrete cyclic distribution due to the cyclostationary nature at
the cyclic frequency axis for $\alpha$.

In the second example, the $f_h$ and $f_r$ frequencies are assumed equal to 1.6  and 0.7, respectively,
and the heart and respiration amplitudes equal to 0.45 mm and 7 mm respectively. With normal values of SNR = SINR = --40 dB for the heart signal and SNR = SINR = --20 dB for the respiration signal, the heart and respiration frequencies
and amplitudes obtained are equal to 1.622 Hz, 0.712 Hz, 0.462 mm
and 9.31 mm, respectively.

\section{Experiment Setup}
Fig.5  shows the designed 2.4GHz miniature radar sensor with the size of 5cm $\times$ 5cm. The radar sensor was configured with a ZigBee module for wireless data transmission, which allows wireless monitoring of the heartbeat and respiration of the subject. This CW Doppler radar system is AC-coupled and the configuration of homodyne direct conversion architecture was adopted to design it. The output I/Q signal will be amplified by an operational amplifier (OP) and read out by an NI-DAQ \cite{24}. The three lead Tektronix 412 ECG is used as reference system to validate the measurement errors.

\begin{figure}[!t]
\centering
\includegraphics[width=2in]{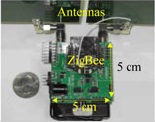}
\caption{The miniature radar system}
\label{fig5}
\end{figure}

We used two of these systems at two corners to cover the space around the subject in both two directions. Fig.6.

\begin{figure}[!t]
\centering
\includegraphics[width=2in]{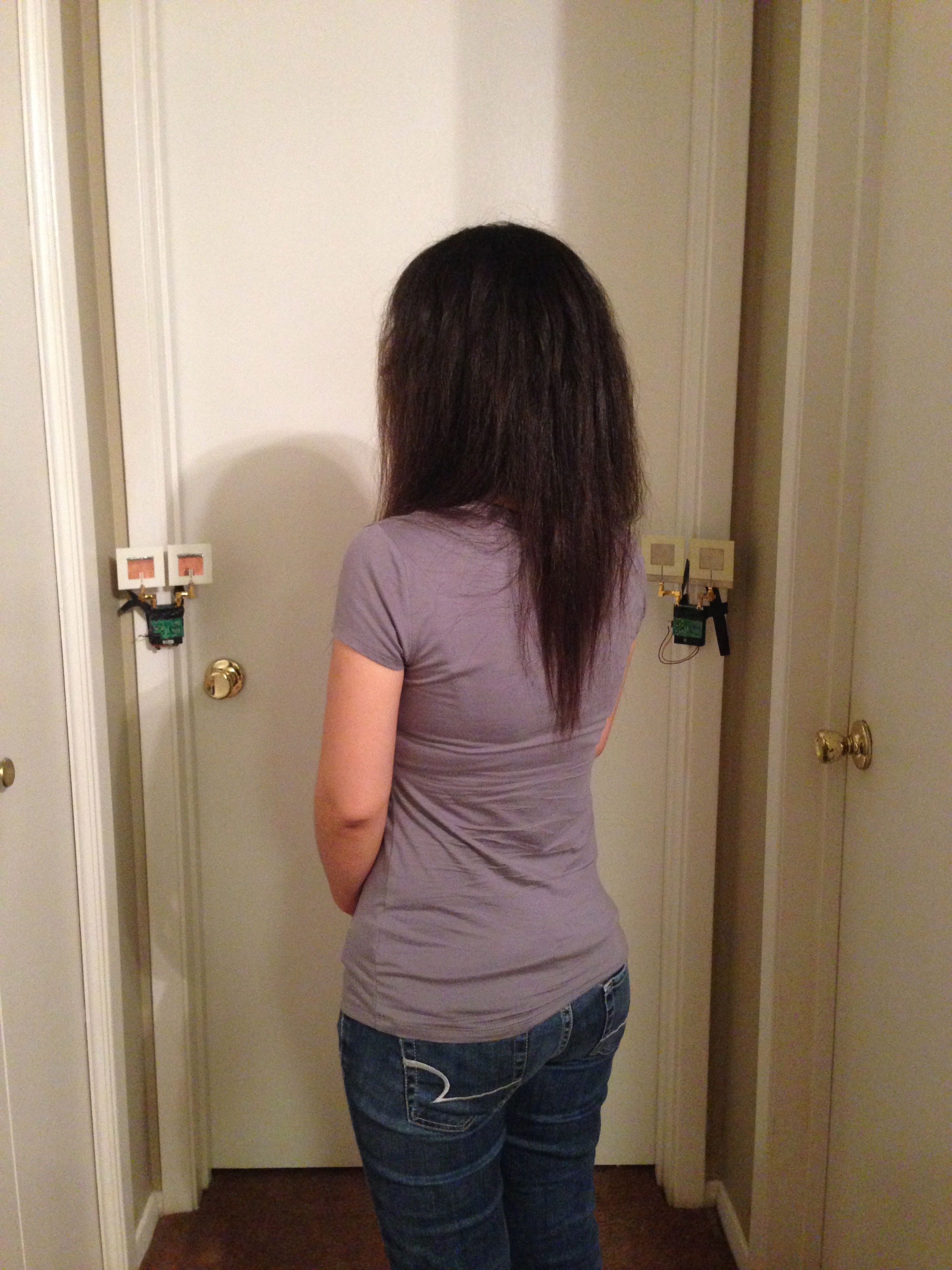}
\caption{Two radar systems at two corners}
\label{fig6}
\end{figure}

At first test, the stationary subject stand at 1m distance from both of two systems and data was recorded every 0.05s during 90s. I and Q recorded signals are shown in Fig.7.

\begin{figure}[!t]
\centering
\includegraphics[width=3.5in]{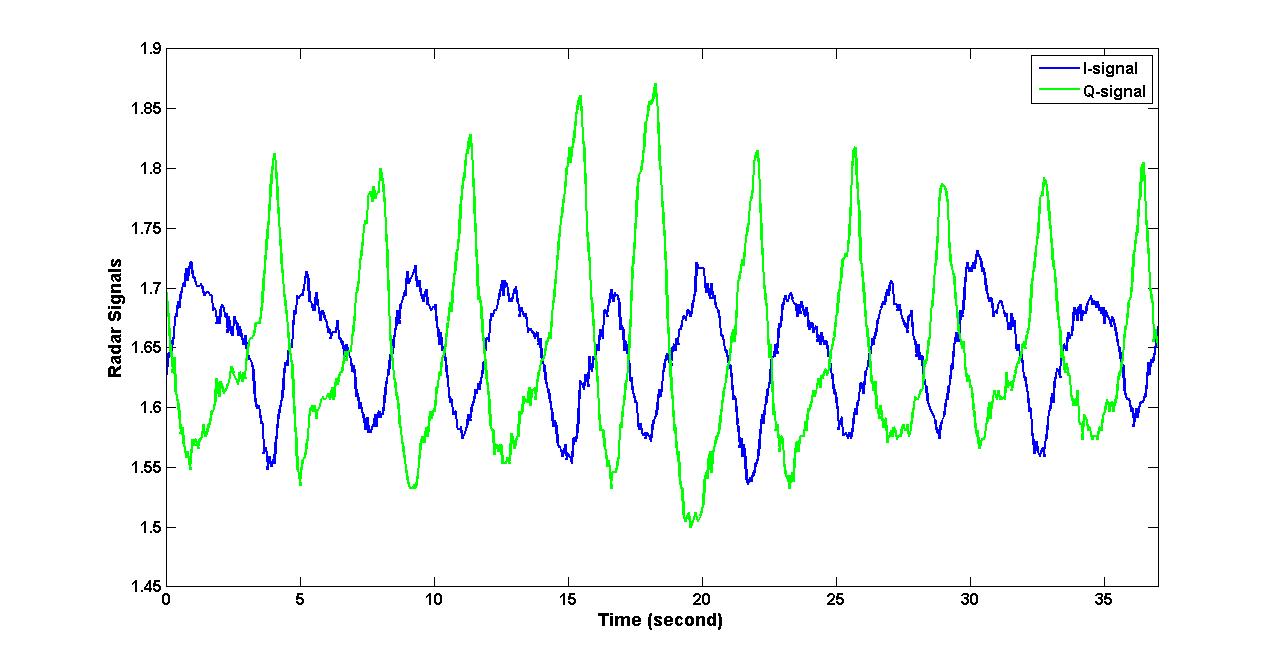}
\caption{I and Q modulated radar signals}
\label{fig7}
\end{figure}

These signals are amplitude and frequency modulated with heart beat and respiration chest wall motions. We apply FFT and cyclostationary algorithms to extract the heart and respiration information from modulated signals. The spectrum and SCF graph of analytical signal $y(t)$ is shown in Fig.8. It can be seen that the heart and respiration peaks could not be shown using FFT analysis but two peaks of heart and respiration are obvious in the SCF graph.

\begin{figure}[!t]
\centering
\includegraphics[width=3in]{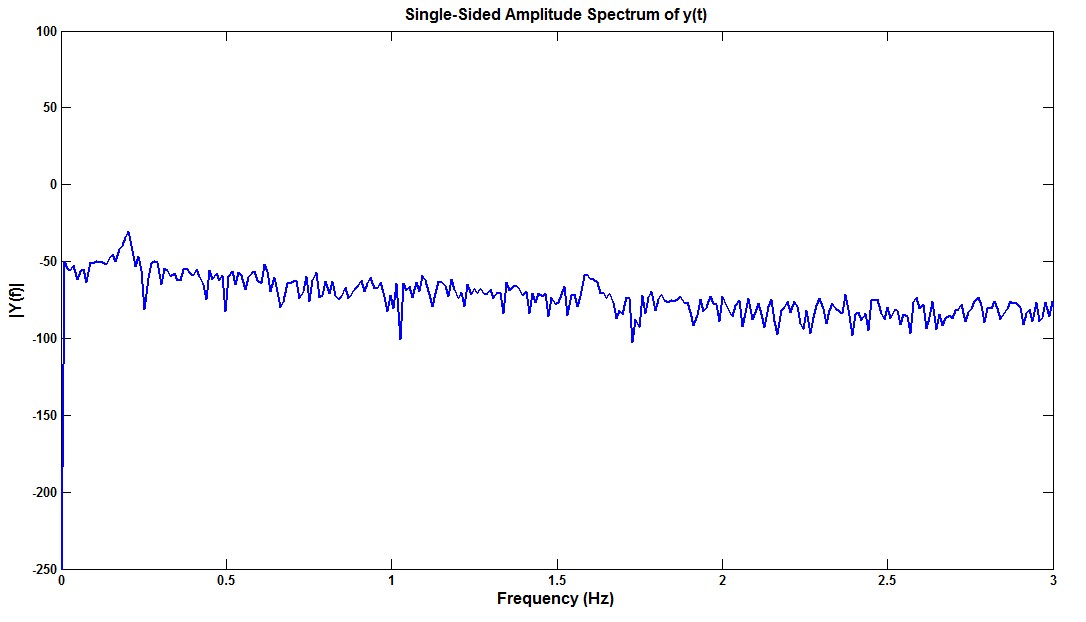}
\begin{center}
(a)
\end{center}
\centering
\includegraphics[width=3in]{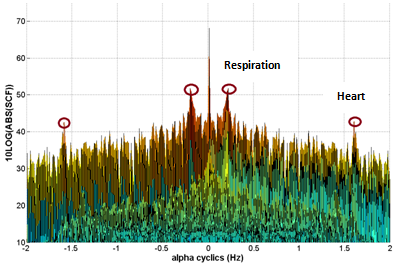}
\begin{center}
(b)
\end{center}
\centering
\caption{ (a) The spectrum and (b) SCF graph of analytical signal $y(t)$}
\label{fig8}
\end{figure}

The second experiment is due the person which has slightly movement in front of the systems. The amounts of body movement dose not proceed from 10 cm in line of sight of the radar system. The effect of movement is addition of noise in spectrum but it is canceled in SCF graph with acceptable level of performance. Fig.9.

\begin{figure}[!t]
\centering
\includegraphics[width=3in]{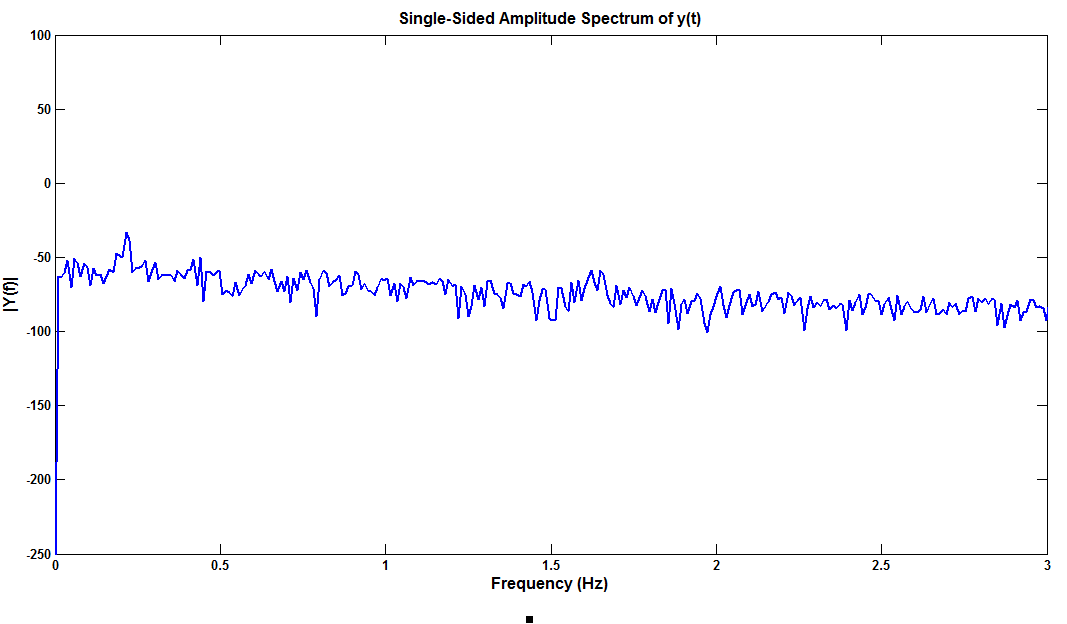}
\begin{center}
(a)
\end{center}
\centering
\includegraphics[width=3in]{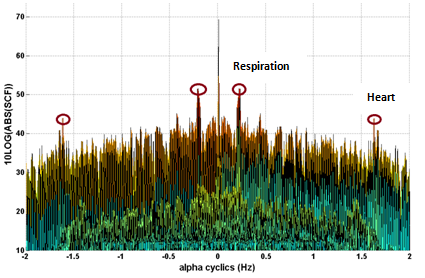}
\begin{center}
(b)
\end{center}
\centering
\caption{ (a)The spectrum and (b) SCF graph of analytical signal $y(t)$}
\label{fig9}
\end{figure}

Fig. 10 show how the root mean square error (RMSE) of the frequency and amplitude estimates varies with the SNR.

\begin{figure}[!t]
\centering
\includegraphics[width=4in]{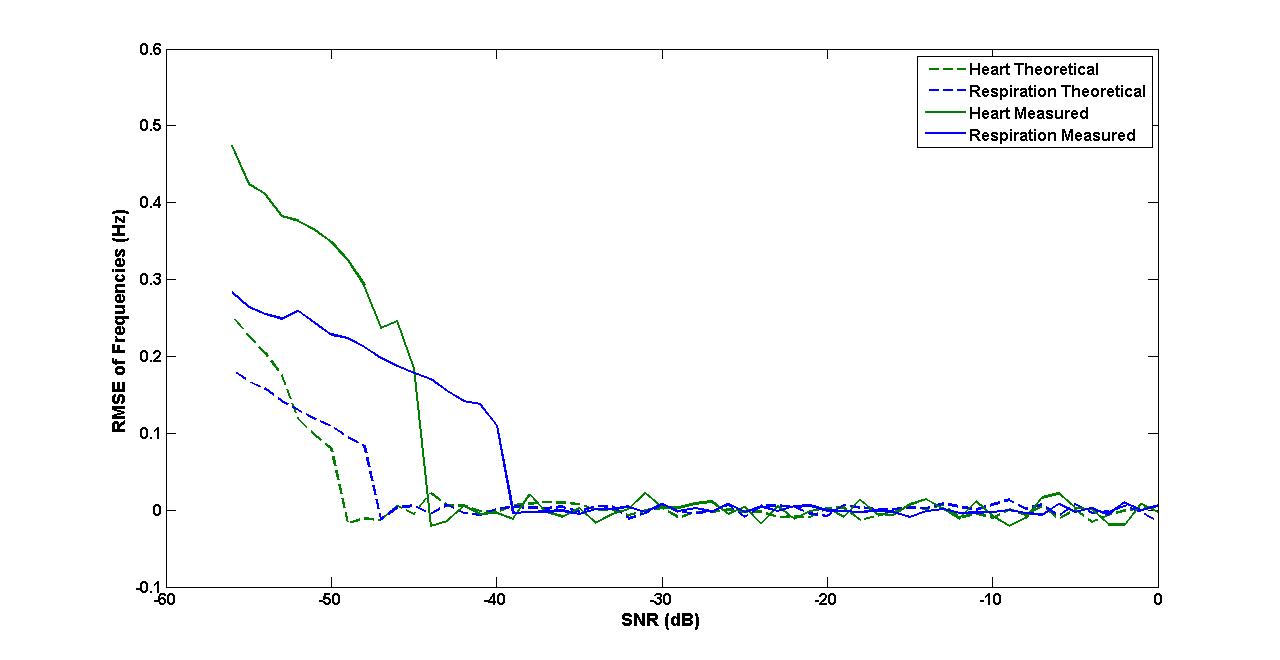}
\begin{center}
(a)
\end{center}
\centering
\includegraphics[width=4in]{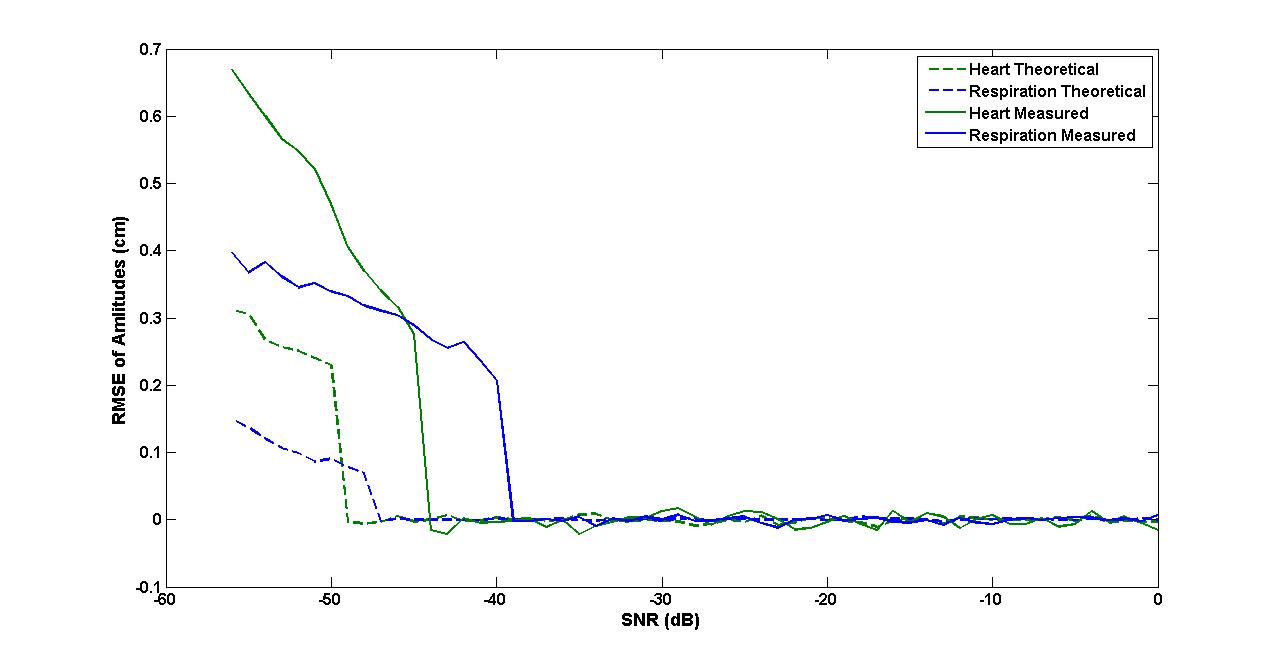}
\begin{center}
(b)
\end{center}
\centering
\caption{Root mean-square error for heart (solid) and respiration (dashed) parameter estimates (a) Frequency. (b) Amplitudes.}
\label{fig10}
\end{figure}
As previously noted, after the threshold has been reached, the simulation results indicate that the
cyclostationary analysis is insensitive to the level of SNR and can accurately
extract vital-sign information even with low SNR and
SINR. At very low SNR the signal will eventually fade in multiplicative noise, so the error then increase.
It is also noted that at very low SNR, the delta function near the offsets $C_{N_\mathrm{I}}$ and $C_{N_\mathrm{Q}}$, defined in (\ref{eq:a19}), would be reduced, so the error is high \cite{25}.

\section{Conclusion}

We have shown that by applying cyclostationary theory to a complex
radar signal, the heart and respiration rates and amplitudes can
be accurately estimated. By the fact that rate detection should be
robust at low SNR such as in environments with high noise, long ranges, and weak signals, or in situations where body motion is large, common signal processing tools and time-frequency approaches fail to extract the periodic
components information, whereas vital sign detection using
cyclostationary theory, would be insensitive to SNR levels, without
any needing phase unwrapping or demodulation.


%

\appendices
\section{}
\label{AppA} In this appendix, we prove that the auto-correlation
function of signal $m(t)$ in (\ref{eq:a11}), is independent of time.
First it is necessary to model phase noise and also random
motion of the subject. There are three main sources of noise in a Doppler radar
system \cite{2}:

\begin{enumerate}
\item     $\phi_\mathrm{r} (t)$: Residual phase noise from the RF oscillator
\item     $\phi_\mathrm{RF}(t)$: RF AWGN
\item     $\phi_\mathrm{f} (t)$: $1/f$ flicker noise from the mixer and baseband circuit
\end{enumerate} These noise sources are uncorrelated and therefore, when they have
been converted to baseband, their powers are additive. The three baseband noise powers are formulated as
follows. The baseband noise power from residual phase noise is expressed
\begin{eqnarray}
N_{\phi_\mathrm{r}(t)}&=&\frac{P_T G^2 G_\mathrm{R} G_\mathrm{C} \sigma_\mathrm{c}}{\pi f^2} S_\phi(1)\nonumber\\
&&\times\ln\left[\frac{f_{\max}}{f_{\min}}\right]\frac{\left(R+\frac{ct_d}{2}\right)}{R^4}.
\end{eqnarray} where $P_T$, $G$, $G_\mathrm{R}$, $G_\mathrm{C}$, $\sigma_\mathrm{c}$ and $S_\phi(1)$ are the transmitted power, antenna gain, receiver gain, mixer conversion gain, radar cross section of the target, and 1-Hz intercept of the phase noise spectrum, respectively. The down-converted RF AWGN noise can be formulated as,
\begin{eqnarray}
N_\mathrm{RF}=8 G_\mathrm{R}G_\mathrm{C}F(kTB)
\end{eqnarray} where $F$, $k$, $T$ and $B$ are, respectively, the noise figure of the receiver, Boltzman's constant, absolute temperature, and bandwidth, Finally, the baseband $1/f$ flicker noise is written
\begin{eqnarray} N_{\phi_\mathrm{f}(t)}=P_{\frac{1}{f}}(1)\ln\left[\frac{f_{\max}}{f_{\min}}\right]
\end{eqnarray} where $P_{\frac{1}{f}}(1)$ is the noise power in a 1-Hz bandwidth centered at 1 Hz.
To prove time-invariance of the autocorrelation function of $m(t)$,
we express,
\begin{eqnarray}
m(t)&=&m_1 (t)m_2 (t)\nonumber\\
\end{eqnarray} where
\begin{eqnarray}
m_1(t)&=&\exp{j\phi_\mathrm{n} (t)}=\exp\{j[\phi_\mathrm{r}(t)+\phi_\mathrm{RF}(t)+\phi_\mathrm{f}(t)]\}\nonumber\\
\end{eqnarray} and
\begin{eqnarray}
m_2(t)&=&\exp\left[j \frac{4\pi}{\lambda} x(t)\right].
\end{eqnarray} According to \cite{34}, the residual phase noise $\phi_r (t)$ for a
free-running oscillator is a non-stationary random process, but
$\exp[j\phi_\mathrm{r}(t)]$ is modeled as stationary. The
$\phi_\mathrm{RF}(t)$ noise is white Gaussian and the $\phi_f (t)$ is
approximated as stationary random noise \cite{2}. By applying the
statistical method for linearization of exponential functions of
random variables \cite{34}, transformation of a random variable $X$ with mean $\mu$
and variance $\sigma^2$ to the random variable $Y=Fe^{AX+B}+G$ can
be approximated by the random variable
$Z=\frac{\sigma_z}{\sigma}(X-\mu)+\mu_\mathrm{z}$, where the $\sigma_z$ and $\mu_z$
parameters can be calculated. Therefore, the exponential function $\exp\{j\left[\phi_\mathrm{RF}(t)+\phi_f(t)\right]\}$
is also approximately stationary, and thus $m_1(t)$ is stationary.

As noted, the Doppler effect resulting from
random motion $x(t)$ in the overall, practical, and least-informative
scenario can be modeled as a 1-D random process, uniformly distributed in the specified interval for both directions.
With this assumption and by applying the linearization method, the
random variable $Y=\exp\left[j\frac{4\pi}{\lambda} x(t)\right]$ also has
uniform distribution. Therefore, $m_2(t)$ is also stationary, and consequently, the autocorrelation function of
$m(t)$ is a function of the lag index only.

\section*{Acknowledgment}
A deep sense of gratitude is expressed to Professor D.R. Morgan for his invaluable help and guidance through some of the issues given in successfully completion of this study. We also appreciate the efforts by Dr. Yang Zhang and Ms. Yiran Li for the hardware setup.

\ifCLASSOPTIONcaptionsoff
  \newpage
\fi


\begin{thebibliography}{1}

\bibitem{1}
M. F. Hilton, R. A. Bates, K. R. Godfrey, M. J. Chappell, and R. M.
Cayton, ``Evaluation of frequency and time-frequency spectral
analysis of heart rate variability as a diagnostic marker of the
sleep apnea syndrome," \emph{Med. Biol. Eng. Comput.}, vol. 37, no. 6, pp.
760-769, 1999.

\bibitem{2}
A. Droitcour, ``Non-contact measurement of heart and respiration
rates with single-chip microwave Doppler radar," Ph.D dissertation,
Dept. Elect. Eng., Stanford University, Stanford, CA, 2006.

\bibitem{3}
K. Chen, J. Zhang, and A. Norman, ``Microwave life-detection systems
for searching human subjects under earthquake rubble or behind
barrier," \emph{IEEE Trans. Biomed. Eng.}, vol. 47, no. 1, pp. 105-114,
Jan. 2000.

\bibitem{4}
S. Bakhtiari, N. Gopalsami, T. W. Elmer, and A. C. Raptis,
``Millimeter wave sensor for far-field standoff vibrometry,"  in \emph{"Proc. AIP Conf.}, vol. 1096 , July 2008, pp. 1641-1648.

\bibitem{5}
V. M. Lubecke, O. Boric-Lubecke, A. Host-Madsen, and A. E. Fathy,
``Through-the-wall radar life detection and monitoring,"  in \emph{IEEE MTT-S Int. Microw. Symp. Dig.}, May 2007, pp. 769-772.

\bibitem{6}
S. Bakhtiari, S. Liao, T. Elmer, N. Sami Gopalsami, and A. C.
Raptis, ``A real-time heart rate analysis for a remote millimeter
wave I-Q sensor,"  \emph{Trans. Biomedi. Eng.}, pp. 1839 - 1845, vol.58, no.6, June 2011.

\bibitem{7}
C. Li and J. Lin, ``Random body movement cancellation in Doppler radar
vital sign detection," \emph{IEEE Trans. Microw. Theory Tech.}, vol. 56, no. 12, pp. 3143-3152, Dec. 2008.

\bibitem{8}
B.-K. Park, O. Boric-Lubecke, and V. M. Lubecke, ``Arctangent
demodulation with dc offset compensation in quadrature Doppler radar
receiver systems," \emph{IEEE Trans. Microw. Theory Tech.}, vol. 55, no. 5,
pp. 1073-1079, May 2007.

\bibitem{9}
H. J. Kim, K. H. Kim, Y. S. Hong, and J. J. Choi, ``Measurement of
human heartbeat and respiration signals using phase detection
radar," \emph{Rev .Sci. Instrum.}, vol.  78, no. 10, pp. 104703-104703-3, Oct 2007.

\bibitem{10}
F. E Churchill, G. W. Ogar, and B. J, Thompson, ``The correction of I and
Q errors in a coherent processor," \emph{IEEE Trans.  Aerosp. Electron. Syst.}, vol. AES-17, no. 1, pp 131-137, Jan. 1981.

\bibitem{11}
D. G. Brennan, ``Linear diversity combining techniques,"  \emph{Proc.IEEE}, vol. 91, pp. 331-356, Feb. 2003.


\bibitem{12}
O. Boric-Lubecke, V. M. Lubecke, I. Mostafanezhad1, B.Kwon Park, W.
Massagram, and B. Jokanovic, ``Doppler radar architectures and signal processing for heart rate extraction,"   \emph{Microw. Rev.}, vol. 15, no. 2, pp. 12-17, 2009.

\bibitem{13}
B.-K. Park, V. Lubecke, O. Boric-Lubecke, and A. Host-Madsen,
``Center tracking quadrature demodulation for Doppler radar motion
detector," in \emph{Proc. IEEE/MTT-S Int. Microwave Symposium}, Jun. 2007, pp. 1323- 1326.
\bibitem{14}
N. Chernov, and C. Lesort, ``Least squares fitting of circles", \emph{Math. Imaging  Vision}, vol. 23, no. 3, pp.
239-252, Nov. 2005.

\bibitem{15}
 M. Zakrzewski, H. Raittinen, and J. Vanhala, ``Comparison of center estimation algorithms for heart and respiration monitoring with
microwave Doppler radar,"  \emph{IEEE Sensors}, vol. 12, no. 3, pp. 627-634, Mar. 2012.

\bibitem{16}
A. Host-Madsen, N. Petrochilos, O. Boric-Lubecke, V. M. Lubecke, B.
K. Park, and Q. Zhou, ``Signal processing methods for Doppler radar
heart rate monitoring,"  in D. Mandic et al (Eds): \emph{Signal Processing
Techniques for Knowledge Extraction and Information Fusion.} Springer-Verlag, Berlin, 2008.

\bibitem{17}
W. Massagram, N. Hafner, B.-K. Park, O. Boric-Lubecke, A.
Host-Madsen, and V. Lubecke, ``Feasibility of heart rate variability
measurement from quadrature Doppler radar using arctangent
demodulation with DC offset compensation,"  in \emph{Proc. 29th IEEE Ann. Int. Conf Eng. In Medicine and Biology Society}, Aug. 2007, pp. 1643-1646.

\bibitem{18}
W. Massagram, V. Lubecke, A. Host-Madsen, and O. Boric-Lubecke,
``Assessment of heart rate variability and respiratory sinus
arrhythmia via Doppler radar," \emph{IEEE Trans.  Microw. Theory
Tech.}, vol. 57, no. 10, pp. 2542-2549, Oct. 2009.


\bibitem{19}
C. Li, and J. Lin, ``Complex signal demodulation and random body
movement cancellation techniques for non-contact vital sign
detection," in \emph{Proc. IEEE/MTT-S Int. Microwave Symposium}, Jun. 2008, pp. 567- 570.

\bibitem{20}
D. R. Morgan and M. G. Zierdt, ``Novel signal processing techniques for
Doppler radar cardiopulmonary sensing," \emph{Signal Processing}, vol. 89,
no. 1, pp. 45-66, Jan. 2009.

\bibitem{21}
W. A. Gardner, \emph{Statistical Spectral Analysis: A Nonprobabilistic
Theory}, Prentice-Hall, Englewood Cliffs, NJ, 1987.

\bibitem{22}
G .B. Giannakis, "Cyclostationary signal analysis" in: V.K.
Madisetti, D.B. Williams (Eds.), \emph{The Digital Signal Processing
Handbook}, CRC Press. and IEEE Press., Boca Raton, FL and New York,
1998, Chap. 17.

\bibitem{23}
M. Derakhshani, ``Efficient cooperative cyclostationary spectrum
sensing in cognitive radios at low snr regimes," \emph{IEEE Trans. Wireless Commun.}, vol. 10, no. 11, pp. 3754-3764, Nov. 2011.

\bibitem{24}
C. Gu, R. Li, H. Zhang,A. Y. C. Fung, C. Torres, S. B. Jiang, C.Li, "Accurate respiration measurement using DC-coupled continuous-wave radar sensor for motion-adaptive cancer radiotherapy," \emph{IEEE Trans Biomed Eng.}, vol. 59. no. 11, pp. 3117-3123, Nov. 2012.

\bibitem{25}
 H. Sadeghi, ``Cyclostationarity-based cooperative spectrum
sensing for cognitive radio networks," in \emph{Proc. International Symposium on Telecommunications} (IST), 27-28 Aug. 2008, pp. 429-434.

\bibitem{26}
Z. Quan, ``Optimal spectral feature detection for spectrum sensing
at very low  SNR," \emph{IEEE Trans. Commun.}, vol. 59, no. 1, pp. 201-212, Jan. 2011.

\bibitem{27}
L. R. Rabiner, M. J. Chang, A. E. Rosenberg, and C. A. McGonegal, ``A
comparative performance study of several pitch detection
algorithms," \emph{IEEE Trans. Acoust., Speech, Signal Process.}, vol. ASSP-24, no. 5, pp. 399-418, Oct. 1976.

\bibitem{28}
B. G. Quinn and E. J. Hannan,  \emph{The Estimation and Tracking of
Frequency}, Cambridge, U.K.: Cambridge Univ. Press., 2001.

\bibitem{29}
P. D. Welch, ``The use of fast Fourier transform for the estimation
of power spectra: A method based on time averaging over short,
modified periodograms," \emph{IEEE Trans. Audio Electroacoust.}, vol.
AU-15, pp. 70-73, Jun. 1967.

\bibitem{30}
R. O. Schmidt, ``Multiple emitter location and signal parameter
estimation,"  \emph{IEEE Trans. Antennas Propagat.}, vol. AP-34, pp.
276-290, Mar. 1986.

\bibitem{31}
B. P. Lathi, \emph{Modern Digital and Analog Communication Systems}, Oxford University Press., 1998.

\bibitem{32}
W. A. Gardner. ``The  spectral correlation theory of cyclostationary
time-series," \emph{IEEE Trans., Commun.}, vol. 11, pp. 13–36, Jul. 1986.

\bibitem{33}
E. Perez Serna, S. Thombre, M. Valkama, S. Lohan, V. Syrj�l�, M.
Detratti, H. Hurskainen, aand J. Nurmi, "Local oscillator phase noise
effects on GNSS code tracking," \emph{Inside GNSS}, pp. 52-62, vol. Nov./Dec.
2010.

\bibitem{34}
J. Farison, ``Approximation of exponential functions of random
variables by statistical linearization," \emph{IEEE Trans. Autom. Control}, vol. 13, pp. 174-178,  Apr. 1968.


%
%
%
%
%
%
%
%
%
%
%
%
%
%
%
%
%
%
%
%
%
%
%
%
%
%
%

\end{thebibliography}
\end{document}